\documentclass[twocolumn,aps,superscriptaddress,showpacs,nofootinbib,floatfix]{revtex4}

\usepackage{epsfig,bm,feynmf}

\usepackage{graphics}

\usepackage[normalem]{ulem}  
\usepackage[dvips]{color} 

\begin{document}


\title{The flow of heavy flavor in hydrodynamics}


\author{Taesoo Song}\email{songtsoo@yonsei.ac.kr}
\affiliation{Cyclotron Institute, Texas A$\&$M University, College Station, TX 77843-3366, USA}
\author{Woosung Park}\email{diracdelta@hanmail.net}
\affiliation{Institute of Physics and Applied Physics, Yonsei
University, Seoul 120-749, Korea}
\author{Su Houng Lee}\email{suhoung@yonsei.ac.kr}
\affiliation{Institute of Physics and Applied Physics, Yonsei
University, Seoul 120-749, Korea}


\begin{abstract}
The flow of charm is calculated in 2+1 ideal hydrodynamics by introducing the charge of $c\bar{c}$ pair assuming that the number of $c\bar{c}$ pairs is conserved in relativistic heavy-ion collisions.
It is found that the mean radial flow velocity of charm quarks is smaller than that of bulk matter by 10$\sim$15 \% and the measured $v_2$ of heavy-flavor electrons is reproduced up to $p_T^e=$ 1.5 GeV/c in Au+Au collision at RHIC.
The same flow is applied to regenerated $J/\psi$ and its $v_2$ is discussed.
\end{abstract}

 \keywords{}

\maketitle

\section{introduction}

Heavy flavor is one of the important probes for the hot dense nuclear matter created in relativistic heavy-ion collisions \cite{Matsui:1986dk,Vogt:1999cu}.
Unlike the light flavors, heavy flavor was expected to be much less thermalized in the matter due to the relatively smaller cross section.
However, it was found that $R_{AA}$ of heavy flavor is much suppressed in the intermediate and high transverse momentum region and that its $v_2$ is not small \cite{Adare:2006nq,Adare:2010de}.
This implies that charm quarks, which comprise  the majority of heavy flavors produced in heavy-ion collisions, are considerably thermalized.
On the other hand, charm quarks never reach chemical equilibrium due to too small cross section for the annihilation or the creation of the $c\bar{c}$ pair.
In fact, the measured number of $c\bar{c}$ pairs in Au+Au collision at RHIC is almost proportional to the number of binary collisions \cite{Adare:2010de}.

Hydrodynamics, which is based on conservation of the energy-momentum and various kind of charges, was found to be successful in reproducing the $p_T$ spectrum and the elliptic flow of bulk particles in relativistic heavy-ion collisions \cite{Heinz:2004qz,Song:2009gc}.  Additionally,
from the fact that the number of charm pairs is almost conserved in the fireball \cite{Adare:2010de}, one can introduce the charge of charm pairs.
This charge is different from the charge of charm flavor, which is zero throughout the space-time of the collision as the the same number of anticharm quarks cancels the charm charge. Once the charge is introduced, its transport can be described in hydrodynamics assuming that the charm quark interact strongly with the matter.

The flow velocity of heavy flavor is a key ingredient in the $v_2$ of open charms as well as that of the charmonia.
In most cases, the flow velocity is assumed same as that of the bulk particles or is just parameterized according to a given assumption \cite{Zhao:2007hh,Akkelin:2009nx,Song:2010er}.
In this study, we attempt to calculate the flow velocity of charm quarks in hydrodynamics and use it to obtain the $v_2$ of the open as well as hidden charm.

The paper is organized as follows: In Sec. \ref{hydro}, our 2+1 hydrodynamics simulation is presented and tested by comparing the calculated $v_2$ of light hadrons with the experimental data at RHIC.
We introduce the charge of charm pairs into the hydrodynamics in Sec. \ref{applications}, and use it to calculate the radial and elliptic flows of charm.
In Sec.~\ref{jpsi}, the elliptic flow of regenerated $J/\psi$ is obtained with the charm flow.
The summary is given in Sec.~\ref{discussion} and the details on the semileptonic decay of open charm in the Appendix.

\section{2+1 hydrodynamics for bulk particles}\label{hydro}

For our 2+1 hydrodynamics, we use the $(\tau,x,y,\eta)$ coordinate system defined as follows,
\begin{eqnarray}
\tau&=&\sqrt{t^2-z^2}, ~~~~\eta=\frac{1}{2}\ln \frac{t+z}{t-z}.
\end{eqnarray}

After relativistic heavy-ion collision, hot dense nuclear matter is produced between the two receding nuclei.
The matter is assumed to be thermalized at $\tau_0=$ 0.6 fm/c \cite{Hirano:2001eu}.
The local entropy density at initial time is parameterized by
\begin{eqnarray}
\frac{ds}{d\eta}=A\bigg\{(1-\alpha)\frac{n_{\rm part}}{2}+\alpha~n_{\rm coll}\bigg\},
\label{entropy}
\end{eqnarray}
where $n_{\rm part(coll)}$ is the number density of participants (binary collisions) in the Glauber model, defined by
$n_{\rm part(coll)}\equiv dN_{\rm part(coll)}/(\tau_0dxdy)$;
The parameters $A$ and $\alpha$ are, respectively, taken as 25.5 and 0.11 to reproduce the multiplicity of charged particles at RHIC \cite{Song:2011xi,Kharzeev:2000ph}.

After the initial thermalization, the hot dense nuclear matter is assumed to expand according to the hydrodynamics.
The equations of energy-momentum conservation in 2+1 hydrodynamics are given as \cite{Teaney:2001av,Heinz:2005bw}
\begin{eqnarray}
\partial_\tau (\tau T^{00})+\partial_x (\tau T^{0x})+\partial_y (\tau T^{0y})&=&-p,\nonumber\\
\partial_\tau (\tau T^{0x})+\partial_x (\tau T^{xx})+\partial_y (\tau T^{xy})&=&0,\nonumber\\
\partial_\tau (\tau T^{0y})+\partial_x (\tau T^{xy})+\partial_y (\tau T^{yy})&=&0,
\label{conservations}
\end{eqnarray}
where $T^{\mu \nu}=(e+p)u^\mu u^\nu-p~g^{\mu\nu}$ with $u_\mu$, $e$ and $p$ being respectively the four-velocity of flow, energy density and pressure of the matter.

\begin{figure}[h]
\centerline{
\includegraphics[width=8.5cm]{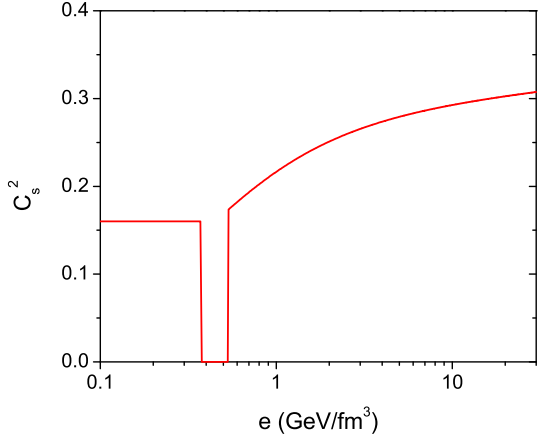}}
\caption{the speed of sound squared as a function of energy density}
\label{sound}
\end{figure}

For the equations of state of QGP and hadron gas, the quasiparticle model based on lattice data and the resonance gas model are adopted respectively \cite{Levai:1997yx,Song:2010ix}.
Whence, a first-order phase transition is assumed and the critical temperature $T_c$ is 170 MeV.
Fig. \ref{sound} shows the speed of sound squared, $c_s^2=\partial p/\partial e$, as a function of energy density.
$c_s^2$ approaches 1/3 at high energy density, which is the value for a free gas, and is zero in the mixed phase where pressure is constant in the first-order phase transition.

The equations in Eq. (\ref{conservations}) are solved numerically by using HLLE algorithm (Godunov-type algorithm) \cite{Schneider:1993gd,Rischke:1995ir,Rishke:1998}.
The simulation is monitored by entropy conservation condition \cite{Teaney:2001av},
\begin{eqnarray}
\frac{dS_{\rm tot}}{d\eta}=\int dxdy\tau s\gamma_\bot,
\label{entropy-2}
\end{eqnarray}
where $s$ is entropy density and $\gamma_\bot=(1-v_x^2-v_y^2)^{-1/2}$.
It is found that the entropy in our simulation is conserved within $\pm$ 4\%  error.

\begin{figure}[h]
\centerline{
\includegraphics[width=8.5cm]{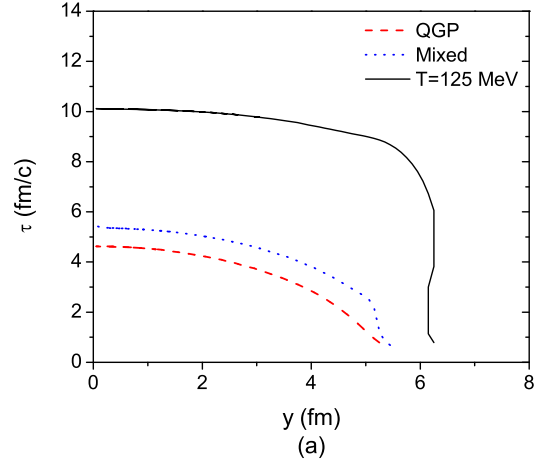}}
\centerline{
\includegraphics[width=8.5cm]{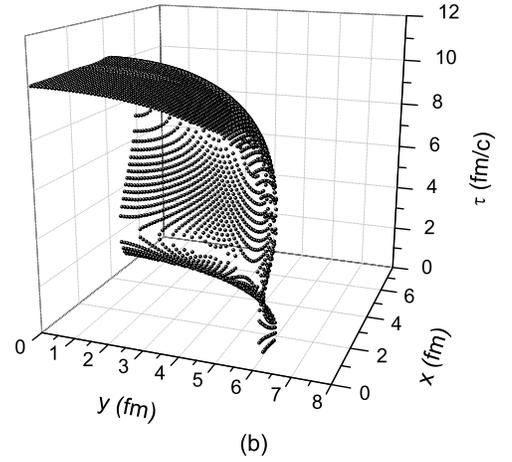}}
\caption{(a) isothermal lines in $(y,\tau)$ plane, where $y$ is perpendicular to both the beam direction and the direction of impact parameter, and (b) $\tau_f(x,y)$ at $b=$ 9 fm in Au+Au collision at RHIC.}
\label{hypersurface}
\end{figure}

Fig. \ref{hypersurface} (a) shows isothermal lines at the impact parameter $b=$ 9 fm.
The temperature for kinetic freeze-out is assumed to be at 125 MeV.
The lifetimes of the QGP, the Mixed and the HG phases are respectively 4.6, 5.4 and 10.1 fm/c.

The differential yield of particles of type $i$ in Cooper-Frye freeze-out formula is
\begin{eqnarray}
\frac{dN_i}{dy m_Tdm_T d\varphi_p}=\frac{1}{(2\pi)^3}\int_{\Sigma_f} p\cdot d^3\sigma(x) f_i(x,p),
\label{dn1}
\end{eqnarray}
where $\varphi_p$ is the azimuthal angle of momentum $p$; $\Sigma_f$ is the hypersurface at freeze-out temperature; $f_i(x,p)$ is the phase-space distribution of particles.
$p^\mu$ and $d^3\sigma_\nu$ in $(\tau,x,y,\eta)$ coordinate system are, respectively,
\begin{eqnarray}
p^\mu=\bigg(m_T\cosh(y-\eta),~p_x,~p_y,~\frac{m_T}{\tau}\sinh(y-\eta)\bigg),\nonumber\\
d^3\sigma_\nu=\bigg(1, -\partial_x \tau_f, -\partial_y \tau_f,~0\bigg)\tau_f(x,y)d\eta dxdy,~~~~~~
\label{dsig}
\end{eqnarray}
where $\tau_f(x,y)$ is the invariant time $\tau$ at freeze-out temperature on the  $(x,y)$ plane, such as shown in Fig. \ref{hypersurface} (b).
Substituting Eq. (\ref{dsig}) into Eq. (\ref{dn1}) \cite{Heinz:2004qz},

\begin{eqnarray}
\frac{dN_i}{dy m_Tdm_T d\varphi_p}=\frac{2g_i}{(2\pi)^3}\sum_{n=1}^{\infty}(\mp)^{n+1}\int dx dy e^{n u_\bot\cdot p_T/T}\nonumber\\
\times\tau_f\bigg[m_T K_1\bigg(\frac{n m_T\gamma_\bot}{T}\bigg)-p_T\cdot \nabla_\bot \tau_f ~K_0 \bigg(\frac{n m_T\gamma_\bot}{T}\bigg)\bigg],
\label{dn2}
\end{eqnarray}
where $g_i$ is degeneracy factor. The sign behind summation is positive for boson and negative for fermion, $K_{0(1)}$ is modified Bessel function, and $\nabla_\bot=(\partial_x, \partial_y)$. If $\tau_f$ does not depend on the transverse position $(x,y)$, Eq. (\ref{dn2}) is equivalent to the blast wave model.

Two comments are to be made in regard to  Eq. (\ref{dn2}). First, if temperature at $(\tau_f+\varepsilon,x,y)$ is higher than $(\tau_f-\varepsilon,x,y)$ with infinitesimal positive $\varepsilon$, the square bracket in Eq. (\ref{dn2}) should be multiplied by an additional negative sign, because the direction of $d^3\sigma_\nu$ is defined as from the higher to lower temperature.
Second, if the square bracket is negative, it is abandoned. Negative $p\cdot d\sigma^3$ means particle moves into the fireball \cite{Rishke:1998}.

\begin{figure}[h]
\centerline{
\includegraphics[width=9 cm]{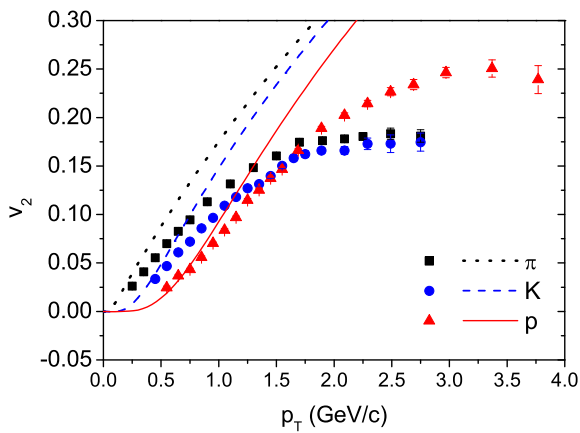}}
\caption{(color online) $v_2$ of $\pi^\pm$, $K^\pm$ and proton at $b=$ 9 fm in hydrodynamics and the experimental data in 20$\sim$60 \% central collision of Au+Au at RHIC \cite{Afanasiev:2007tv}.}
\label{v2-fig}
\end{figure}

The $v_2$ of light hadrons from our hydrodynamics are compared with experimental data in Fig. \ref{v2-fig}. Here,  $v_2$ of particle $i$ is defined as the following:

\begin{eqnarray}
v_2(p_T)=\frac{\int d\varphi_p \cos(2\varphi_p) dN_i/(dy m_T dm_T d\varphi_p)}{\int d\varphi_p dN_i/(dy m_T dm_T d\varphi_p)}.
\label{v2}
\end{eqnarray}

The experimental data include the decay of particles like $\rho \rightarrow \pi+\pi$, but ours do not.
As in other hydrodynamics simulations, $v_2$ is reproduced well at low $p_T$ including the ordering of $v_2$ according to the particle masses.

\section{2+1 hydrodynamics for heavy flavor}\label{applications}

The charge conservation in 2+1 hydrodynamics is expressed as

\begin{eqnarray}
\partial_\tau (\tau n)+\partial_x (\tau nv_x)+\partial_y (\tau nv_y)=0,
\label{charge}
\end{eqnarray}
where $n$ can be any kind of charge density that is conserved.
The number of $c \bar{c}$ pairs produced in relativistic heavy-ion collisions, which is proportional to the number of binary collisions, is assumed to be conserved during the fireball expansion.  The justification for this assumption is the small cross sections for the creation or the annihilation of the pair and the overall small number densities for these heavy particles \cite{Andronic:2006ky,Adare:2010de}.
Therefore, we can define the charge of $c \bar{c}$ pair.
This charge is different from the charge of charm flavor, because the charm net charge is always canceled by the same number of the anticharm quarks.
Now, assuming (anti)charm quarks interact with nuclear matter as strongly as light quarks, charm quarks follow the bulk flow and Eq. (\ref{charge}) can be used for the transport of $c\bar{c}$ pairs in the nuclear matter.
We further assume that charm has thermal distribution in each fluid cell, though it may fail in high-$p_T$ region.
More realistic description is possible by solving the Langevin equation \cite{Moore:2004tg,vanHees:2005wb}.

The initial charge density of $c\bar{c}$ is given by Glauber model as following \cite{Song:2010ix}:

\begin{eqnarray}
n(\tau_0,x,y)=~~~~~~~~~~~~~~~~~~~~~~~~~~~~~~~~~~~~~~~~~~~~~\nonumber\\
\frac{\sigma_{c\bar{c}}^{pp}AB}{\tau_0} \int_{-\infty}^\infty dz \int_{-\infty}^\infty dz^\prime \rho_A(\vec{s},z) \rho_B(\vec{b}-\vec{s},z^\prime),
\label{initial}
\end{eqnarray}
where $\sigma_{c\bar{c}}^{pp}$ is the cross section for $c\bar{c}$ production in p+p collision; A and B are mass numbers of nucleus A and B respectively; $\vec{s}$ and $\vec{b}$ are the transverse spatial vector and the impact parameter vector respectively; $\rho_{A(B)}$ is the Wood-Saxon distribution function of a nucleon in nucleus A(B).
The transport of heavy flavor is obtained by solving Eq. (\ref{conservations}) and (\ref{charge}) with the initial condition of Eq. (\ref{initial}).

\begin{figure}[h]
\centerline{
\includegraphics[width=8.5cm]{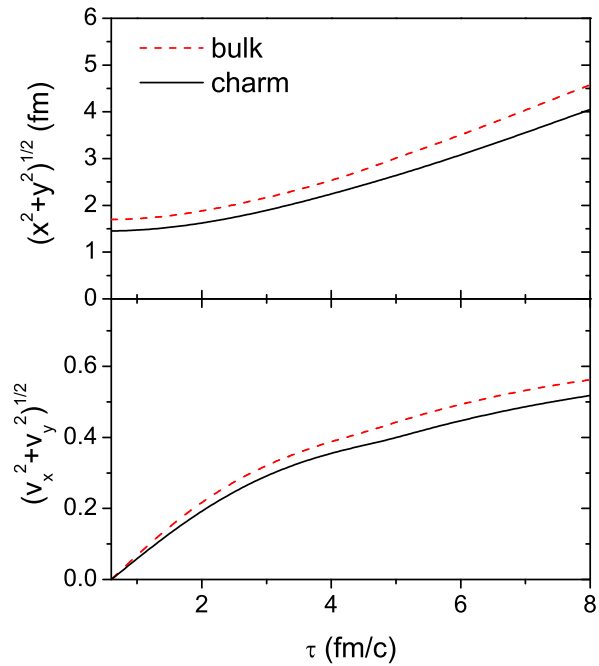}}
\caption{the mean radial distances from the center of fireball and the mean radial flow velocities of bulk matter and of heavy flavor at $b=$ 9 fm in Au+Au collision.}
\label{velocity}
\end{figure}

Fig. \ref{velocity} shows the mean radial distances from the center of fireball and the mean radial flow velocities of the  bulk matter and of the heavy flavor at $b=$ 9 fm in Au+Au collision.
For the bulk matter, the radial distance and the radial flow velocity are weighted by entropy density, and for heavy flavor, by the charge density of heavy flavor, both in the fireball frame.
As a result, the mean radial distance of heavy flavor is smaller than that of the bulk matter by 10$\sim$15 \%.
And the radial flow velocity is also smaller by similar percentages.
The reason is as following:
The number of $c\bar{c}$ pairs is proportional to the number of binary collisions while the multiplicity of bulk particles mainly to the number of participants, as seen in Eq. (\ref{entropy}) and (\ref{initial}).
The number of binary collisions is more concentrated in the central region of fireball than that of the participants.
So the mean radial distance of heavy flavor from the center is smaller than that of bulk matter.
On the other hand, the flow velocity is inversely proportional to the distance from the center, because pressure gradient is high at the border of fireball.
Therefore, $c\bar{c}$ pair is less accelerated than bulk particle in hydrodynamics.

The $v_2$ of open charms is the same as Eq. (\ref{v2}) except that the differential yield is multiplied by the local charm fugacity, $\gamma_c(\tau,x,y)$, which is proportional to the charge density of $c\bar{c}$ in local frame, $n(\tau,x,y)/\gamma_\bot(\tau,x,y)$:

\begin{eqnarray}
v_2^D(p_T)=~~~~~~~~~~~~~~~~~~~~~~~~~~~~~~~~~~~~~~~~~~~~~~~~~~~~~~~~\nonumber\\
\frac{\int d\varphi_p \cos(2\varphi_p)\int dx dy \gamma_c(\tau,x,y) ~dN/(dy m_T dm_T d\varphi_p)}{\int d\varphi_p \int dx dy \gamma_c(\tau,x,y) ~dN/(dy m_T dm_T d\varphi_p)},\nonumber\\
\label{v2a}
\end{eqnarray}

Experimentally the elliptic flow of heavy flavor is measured through the electrons produced from the semileptonic decay of heavy flavor.
To reproduce the experimental data, D mesons are generated according to Eq. (\ref{dn2}) with the local charm fugacity being multiplied.
It is assumed that open charms freeze out at $T_c$.
And then 61.4 \% and 38.6 \% of them decay into $K+e^-+\bar{\nu}_e$ and $K^*+e^-+\bar{\nu}_e$ respectively, assuming that the branching ratio of semileptonic decay is 100 \%.
The energy spectra of single electrons in both decay channels are presented in Appendix.

\begin{figure}[h]
\centerline{
\includegraphics[width=8.5cm]{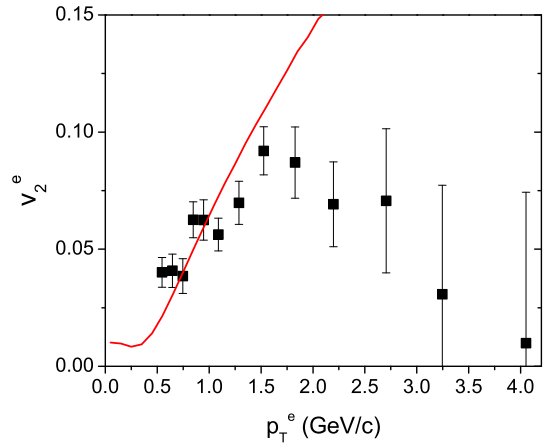}}
\caption{the $v_2$ of heavy-flavor electrons at $b=$ 5.1 fm in hydrodynamics and the experimental data in minimum bias \cite{Adare:2006nq}.}
\label{v2e}
\end{figure}

Fig. \ref{v2e} compares the $v_2$ of heavy-flavor electrons at $b=$ 5.1 fm in hydrodynamics and the experimental data in minimum bias \cite{Adare:2006nq}.
The impact parameter corresponds to the mid-centrality weighted by the number of $c\bar{c}$ pairs \cite{Miller:2007ri}.
It is seen that the $v_2$ in hydrodynamics reproduces the experimental data up to around $p_T^e=$ 1.5 GeV.
This result implies that the heavy flavor in heavy-ion collision at RHIC is considerably thermalized at low transverse momentum.

\section{the elliptic flow of $J/\psi$}\label{jpsi}

The $v_2$ of $J/\psi$ in relativistic heavy-ion collisions is a very interesting quantity revealing how much fraction of $J/\psi$ is produced through regeneration \cite{Song:2010er}.
Ignoring the $v_2$ of primordial $J/\psi$ which seems much smaller than that of regenerated $J/\psi$, the $v_2$ of total $J/\psi$  can be approximated to be  the $v_2$ of regenerated $J/\psi$ multiplied by its fraction among the total observed yield of $J/\psi$.
The fraction gives valuable information on the thermal properties of $J/\psi$ as well as of the hot nuclear matter created in relativistic heavy-ion collision.
Because the $J/\psi$ is composed of charm and anticharm quark, the differential yield of regenerated $J/\psi$ in $v_2$ is multiplied by the local charm fugacity squared:

\begin{eqnarray}
v_2^{J/\psi}(p_T)=~~~~~~~~~~~~~~~~~~~~~~~~~~~~~~~~~~~~~~~~~~~~~~~~~~~~~~~\nonumber\\
\frac{\int d\varphi_p \cos(2\varphi_p)\int dx dy \gamma_c^2(\tau,x,y) ~dN/(dy m_T dm_T d\varphi_p)}{\int d\varphi_p \int dx dy \gamma_c^2(\tau,x,y) ~dN/(dy m_T dm_T d\varphi_p)}.\nonumber\\
\label{v2b}
\end{eqnarray}

\begin{figure}[h]
\centerline{
\includegraphics[width=9 cm]{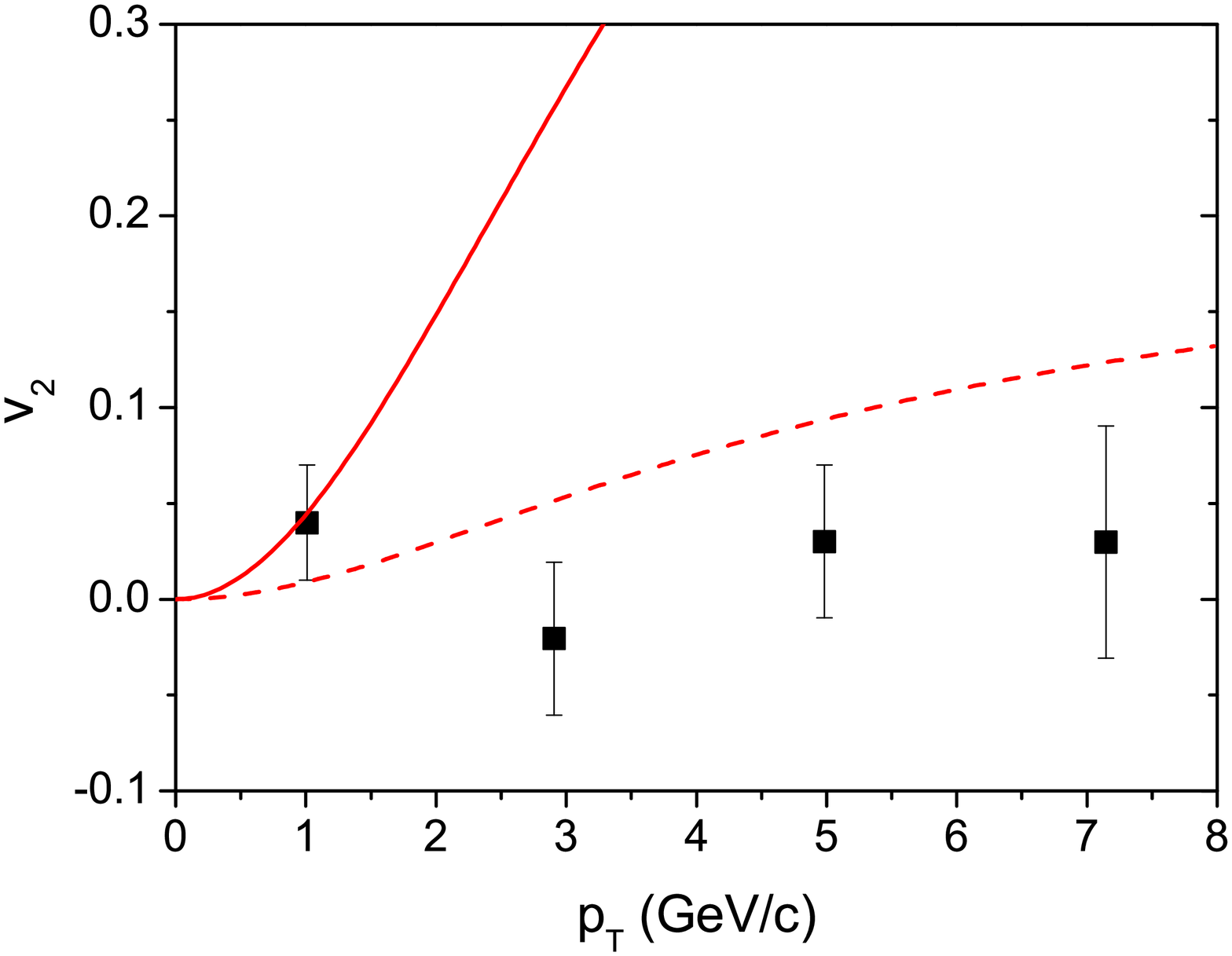}}
\caption{the $v_2$ of regenerated $J/\psi$ at $b=$ 9 fm in hydrodynamics (solid line) and 20 \% of them (dashed line). The experimental data are taken from \cite{Silva}.}
\label{v2jpsi}
\end{figure}

In Fig. \ref{v2jpsi}, the solid line is the $v_2$ of regenerated $J/\psi$ at $b=$ 9 fm, which corresponds to 20$\sim$60 \% central collision.
In contrast to our previous study \cite{Song:2010er}, where only radial component of flow was considered, negative $v_2$ is not observed at low $p_T$.
The freeze-out temperature for $J/\psi$ is assumed to be $T_c$ as in the  open charm case.
The figure shows that we reproduce the first data point at $p_T=$ 1.0 GeV, but overestimate the data at three higher $p_T$.
This result could be explained if the measured $J/\psi$ at these higher $p_T$ are mostly primordial ones, which has no elliptic flow.
In the two-component model, the fraction of regenerated $J/\psi$ is only 20 \% at $b=$ 9 fm \cite{Song:2011xi}.
The dashed line is $v_2$ assuming 20 \% of $J/\psi$ are regenerated ones and the rest primordial ones.
It seems from the figure that the fraction of regenerated $J/\psi$ is larger than 20 \% at $p_T=$ 1.0 GeV, but smaller than 20 \% at three higher $p_T$.

\section{summary}\label{discussion}

The flow of heavy flavor in heavy-ion collision is studied in 2+1 hydrodynamics by introducing the charge of the $c\bar{c}$ pair.
For this, we assume that the number of heavy flavor does not change in fireball but that the heavy flavor interacts with the nuclear matter created in relativistic heavy-ion collision as strongly as the light flavors.

It is found that the mean radial distance from the center of fireball and the mean radial flow velocity of heavy flavor are less than those of bulk particles by 10$\sim$15 \%.
This is so because the creation of heavy flavor is more concentrated in the central region than that of bulk particles.

The elliptic flow of heavy-flavor electrons is calculated in Monte Carlo method and compared with experimental data at RHIC.
If heavy flavor particle does not interact with the nuclear matter, it has no elliptic flow ($v_2=$ 0).
If the interaction is strong enough, heavy flavor particle has the maximum elliptic flow which is given by hydrodynamics.
It is found that our model reproduces experimental data up to around $p_T^e=$ 1.5 GeV.
It implies that the heavy flavors are considerably thermalized at low transverse momentum.

Finally, the $v_2$ of regenerated $J/\psi$ is calculated with the same flow of charm.
By comparing with experimental data, it is found that the fraction of regenerated $J/\psi$ depends on transverse momentum and the the fraction is not small at low $p_T$.

\section*{Acknowledgements}
This work was supported was supported by the Korean Ministry of
Education through the BK21 Program and KRF-2006-C00011.



\appendix
\section{semileptonic decay of open charm}\label{vh}

In this study,  two main semileptonic decay channels of D meson are considered:
$D\rightarrow K+e^-+\bar{\nu}_e$ and $D\rightarrow K^*+e^-+\bar{\nu}_e$ with the average branching ratios being 6.09 \% and 3.84 \% respectively.
In the limit of massless electron, the differential partial widths are
\begin{eqnarray}
\frac{d\Gamma(D\rightarrow K+e^-+\bar{\nu}_e)}{dq^2d\cos\theta}\sim {\bf p}^3|f(q^2)|^2\sin^2\theta,~~~~~~~~~~~~~\label{width1}\\
\frac{d\Gamma(D\rightarrow K^*+e^-+\bar{\nu}_e)}{dq^2d\cos\theta}\sim {\bf p}q^2\bigg\{ \frac{(1-\cos\theta)^2}{2}|H_-(q^2)|^2\nonumber\\
+\frac{(1+\cos\theta)^2}{2}|H_+(q^2)|^2+\sin^2\theta|H_0(q^2)|^2\bigg\},~~~~\label{width2}
\end{eqnarray}
where
\begin{eqnarray}
H^\pm(q^2)=\frac{(m_D+m_{K^*})^2A_1(q^2)\mp 2m_D{\bf p}V(q^2)}{m_D+m_{K^*}},~~~~~~~\nonumber\\
H_0(q^2)=\frac{1}{\sqrt{q^2}}\frac{m_D^2}{2m_{K^*}(m_D+m_{K^*})}~~~~~~~~~~~~~~~~~~~~~~~~\nonumber\\
\times\bigg\{\bigg(1-\frac{m_{K^*}^2-q^2}{m_D^2}\bigg)(m_D^2+m_{K^*}^2)A_1(q^2)-4{\bf p}^2A_2(q^2)\bigg\}.\nonumber
\end{eqnarray}

In Eq. (\ref{width1})$\sim$(\ref{width2}), $q^2$ and ${\bf p}$ are, respectively, the four momentum squared and the magnitude of three momentum of W boson in the D meson rest frame. $\theta$ is the angle between the three momentum of lepton and that of W boson in the W rest frame.
$f(q^2)$, $A_1(q^2)$, $V(q^2)$ and $A_2(q^2)$ are form factors and are parameterized as the followings:

\begin{eqnarray}
F(q^2)=\frac{F(0)}{1-q^2/m_{\rm pole}^2},
\end{eqnarray}
where $m_{\rm pole}=$ 1.89 GeV for $f(q^2)$, and the same value is applied to $A_1(q^2)$, $V(q^2)$ and $A_2(q^2)$ for simplicity. $r_V\equiv V(0)/A_1(0)=1.62\pm 0.08$, and $r_2\equiv A_2(0)/A_1(0)=0.83\pm 0.05$ taken from the particle data book.

\begin{figure}[h]
\centerline{
\includegraphics[width=8.5cm]{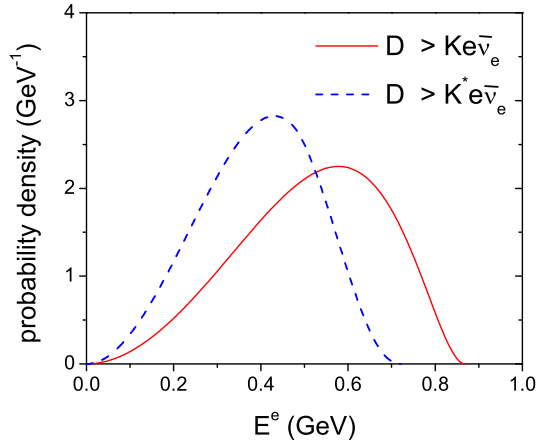}}
\caption{the probability density for the electron in $D\rightarrow K(K^*)+e^-+\bar{\nu}_e$ to have energy $E_e$ in D meson rest frame.}
\label{El}
\end{figure}

In D meson rest frame, the momenta of $K$($K^*$) and W boson are respectively
\begin{eqnarray}
p_\mu=\bigg(\sqrt{m_{K(K^*)}^2+{\bf p}^2},~0,~0,~-{\bf p}~\bigg),\\
q_\mu=\bigg(m_D-\sqrt{m_{K(K^*)}^2+{\bf p}^2},~0,~0,~{\bf p}~\bigg).
\end{eqnarray}

Supposing that the momentum of electron in W boson rest frame
\begin{eqnarray}
p^e_\mu=\bigg(\frac{q}{2},~\frac{q}{2}\sin\theta,~\frac{q}{2}\sin\theta\sin\phi,~\frac{q}{2}\cos\theta\bigg),
\end{eqnarray}

the energy of the electron in D meson rest frame
\begin{eqnarray}
E^e=\frac{q}{2}\gamma(1-\beta\cos\theta)=\frac{1}{2}(q_0+q_z\cos\theta),
\label{leptonE}
\end{eqnarray}
where $\beta=-q_z/q_0$ and $\gamma=q_0/q$, and
\begin{eqnarray}
\cos\theta=\frac{1}{\bf p}\bigg(2E^e-m_D+\sqrt{m_{K(K^*)}^2+{\bf p}^2}\bigg),\\
d\cos\theta=\frac{2}{\bf p}dE^e.~~~~~~~~~~~~~~~~~~~~~~~~~~~~~~~~~~~~~
\end{eqnarray}

Eq. (\ref{width1}) with the new variable, $E^e$, is then
\begin{eqnarray}
\frac{d\Gamma(D\rightarrow K+e^-+\bar{\nu}_e)}{dE^e}\sim \int_0^\alpha dq^2 {\bf p}^3|f(q^2)|^2\sin^2\theta\nonumber\\
\sim-m_{\rm pole}^4 \bigg(1-\frac{2E^e}{m_D}\bigg)\ln \bigg(1-\frac{\alpha}{m_{\rm pole}^2}\bigg)~~~~~~~~~~~~~\nonumber\\
+\bigg[m_{\rm pole}^2\bigg(1-\frac{2E^e}{m_D}\bigg)+4E^{e2}-\frac{m_D^2-m_K^2}{m_D}2E^e\bigg]\nonumber\\
\times\bigg(\frac{m_{\rm pole}^4}{\alpha-m_{\rm pole}^2}+m_{\rm pole}^2\bigg).~~~~
\label{new1}
\end{eqnarray}

where the upper limit of the integration
\begin{eqnarray}
\alpha=2m_D E^e-\frac{2m_K^2E^e}{m_D-2E^e}\nonumber
\end{eqnarray}
is given from the condition that $\sin^2\theta \geq 0$. The available range of $E^e$ is taken to be from 0 to $(m_D-m_K)(1+m_K/m_D)/2$. The phase space is the same for $D\rightarrow K^*+e^-+\bar{\nu}_e$ after $m_K$ is replaced by $m_{K^*}$.

Fig. \ref{El} shows the probability density for an electron from D meson decay to have energy $E^e$ in the D meson rest frame.


\end{document}